\documentclass{pasj00}
\usepackage{color}
\draft

\begin{document}
\SetRunningHead{Y. Takeda \& S. UeNo}{Impact of Hydrogen Collisions 
on the Formation of C~{\sc i} 1.07~$\mu$m Lines}
\Received{2013/09/19}%{yyyy/mm/dd}
\Accepted{2013/10/30}%{yyyy/mm/dd}

\title{Empirical Investigation on the Impact of Hydrogen Collisions for the 
Formation of C~I~1.07~$\mu$m Lines \\
Based on the Solar Center-to-Limb Variation
\thanks{Based on data collected by the Domeless Solar Telescope at 
Hida Observatory (Kyoto University, Japan).}
}

%%% Please use the following style in case that sorting by 
%%% affiliation is impossible. 
%
% \author{%
%   D-Firstname \textsc{D-Familyname}\altaffilmark{1}
%   E-Firstname \textsc{E-Familyname}\altaffilmark{1,2}
%   and
%   F-Firstname \textsc{F-Familyname}\altaffilmark{2}}
% \altaffiltext{1}{Address of Institute}
% \email{ddddd@xxx.xxx.xx.xx}
% \email{eeeee@xxx.xxx.xx.xx}
% \altaffiltext{2}{Address of Institute}

\author{Yoichi \textsc{Takeda}}
\affil{National Astronomical Observatory, 2-21-1 Osawa, 
Mitaka, Tokyo 181-8588}
\email{takeda.yoichi@nao.ac.jp}
\and
\author{Satoru \textsc{UeNo}}
\affil{Kwasan and Hida Observatories, Kyoto University,\\
Kurabashira, Kamitakara, Takayama, Gifu 506-1314}
\email{ueno@kwasan.kyoto-u.ac.jp}

%%% Please use the following style in case that sorting by 
%%% affiliation is impossible. 
%
% \author{%
%   D-Firstname \textsc{D-Familyname}\altaffilmark{1}
%   E-Firstname \textsc{E-Familyname}\altaffilmark{1,2}
%   and
%   F-Firstname \textsc{F-Familyname}\altaffilmark{2}}
% \altaffiltext{1}{Address of Institute}
% \email{ddddd@xxx.xxx.xx.xx}
% \email{eeeee@xxx.xxx.xx.xx}
% \altaffiltext{2}{Address of Institute}

%% `\KeyWords{}' always has to be placed before `\maketitle'.
%\KeyWords{xxxx:xxxx ......} %Do NOT move this preamble from here!

\maketitle

\begin{abstract}
With an aim of examining the validity of non-LTE line-formation calculations
for the strong C~{\sc i} lines of multiplet 1 at 1.068--1.069~$\mu$m,
especially in terms of the treatment of collisions with neutral hydrogen 
(H~{\sc i}) atoms, we computed theoretical equivalent widths ($W_{\lambda}$) 
of these lines corresponding to specific intensities of different angles 
($\mu = \cos\theta$) on the solar disk under various conditions,
which were then compared with the empirical $W_{\lambda}$ vs. $\mu$ relations
obtained from our spectroscopic observations using the Domeless Solar 
Telescope at Hida Observatory.
It turned out that our observational data are almost consistent with 
the theoretical simulations done with the H~{\sc i} collision rates computed
with the classical formula, which suggests that the necessity of its 
significant revision (e.g., considerable reduction) is unlikely.
\end{abstract}

%\section{}
%
%\noindent IMPORTANT NOTICE\\
%1. ``\verb|\draft|'' creates single column and double spaces format.\\
%2. If you comment out ``\verb|\draft|'', the output will be double column
%   and single space.\\
%3. For cross-references, the use of ``\verb|\label|, \verb|\ref|, \verb|\cite|'%' 
%   and the thebibliography environment is strongly recommended. \\
%4. Do NOT use ``\verb|\def|, \verb|\renewcommand|''.\\
%5. Do NOT redefine commands provided by PASJ00.cls.\\

%\newpage

%Sect. 1
\section{Introduction}

Recently, Takeda and Takada-Hidai (2013, hereinafter referred to 
as TTH13) showed that a group of strong C~{\sc i} lines 
at $\sim$~1.07~$\mu$m in the near-IR region 
(multiplet 1; 3s~$^{3}{\rm P}^{\rm o}$ -- 3p $^{3}{\rm D}$;
$\chi_{\rm low} \sim 7.5$~eV) are quite useful for investigating
the carbon abundances of old metal-deficient stars, thanks to their
visibility down to the extremely metal-poor regime irrespective
of the type of stars.
For example, these C~{\sc i} lines are applicable not only to 
metal-poor giants (of lower $T_{\rm eff}$ and lower $\log g$) 
but also to very metal-poor turn-off stars (of higher $T_{\rm eff}$ 
and higher $\log g$), for which CH molecular lines (widely used for 
C abundance determinations of population II stars) become too weak 
to be measurable due to their large sensitivity to $T_{\rm eff}$.
Besides, these near-IR lines in the $J$-band are superior to other 
occasionally-used C~{\sc i} lines of multiplet 3 at 0.91~$\mu$m, 
because of being unaffected by telluric lines.

As TTH13 pointed out, however, these strong near-IR lines suffer
considerably large non-LTE effects ($|\Delta|\sim $~0.1--0.5~dex;
where $\Delta$ is the negative non-LTE abundance correction; 
cf. figure 4b in TTH13), the extents of which tend to progressively 
increase with a decrease in metallicity.
Moreover, these corrections were found to be sensitive to how
the collisional rates due to neutral hydrogen atoms are included; 
i.e., while TTH13 adopted the practical formula derived by
Steenbock and Holweger (1984) based on Drawin's (1968, 1969)
classical cross sections for evaluating these H~{\sc i} collisions, 
$|\Delta|$ would become even doubled if these classical rates 
are reduced by a factor of 10 (cf. figure 4c in TTH13).
This problem has a significant impact in connection with the 
systematic discrepancy between the abundances derived from
these C~{\sc i} lines and CH molecular lines, as discussed in TTH13.
It is thus important to check whether or not the classical
treatment for these H~{\sc i} collisions is reasonable in an 
empirical manner. 

Given this situation, we decided to make use of the center-to-limb 
variation of the strengths of these lines on the solar disk, 
as suggested by TTH13, with an intention to provide some empirical
constraint on this issue by comparing the theoretical results 
(calculated under various conditions) with the new observational data
obtained by ourselves. The purpose of this paper is report on
this investigation.

The remainder of this article is organized as follows. 
After describing our solar observations and measurements 
of equivalent widths in section 2, we explain theoretical
calculations of equivalent widths along with the adopted solar 
model atmospheres in section 3. Section 4 is devoted to 
discussing the characteristics of theoretically computed results, 
which are further compared with the observational data
to reach a conclusion. 
Besides, we briefly mention the results for the 
O~{\sc i} 7771--5~$\rm\AA$ triplet lines (also targeted in our observations)
in appendix~1, followed by appendix~2 where the temperature structure 
of the solar photospheric model adopted in this study is compared 
with those of various other models.

%Sect. 2
\section{Observations and Measurements}

\subsection{Observational Data}

The observations were carried out on 2013 July 26
(JST)\footnote{The aspect angles of the solar rotation axis 
($P$: position angle between the geographic north pole
and the solar rotational north pole; $B_{0}$: heliographic
latitude of the central point of the solar disk) on this date were 
($+8^{\circ}22'$, $5^{\circ}19'$).}
by using the 60~cm Domeless Solar Telescope (DST) with the
Horizontal Spectrograph of the Hida Observatory, Kyoto University
(Nakai \& Hattori 1985).
Regarding the target positions on the solar disk, we selected 12 
points on the meridian line of the solar disk (1 point at the disk center 
and 11 points from 0.45~$R_{0}$ to 0.95~$R_{0}$ with a 
step of 0.05~$R_{0}$), at which the slit was aligned in 
the E--W direction on the Sun as depicted in figure 1.
In the adopted setting of the spectrograph, 
our observation produced a solar spectrum with a resolving power of 
$R \sim 190,000$\footnote{The entrance slit (0.05~mm width) 
projects a shadow corresponding to 0.0288~$\rm\AA$ width on the detector 
while the theoretical resolution of the grating is 0.0468~$\rm\AA$. 
The root-sum-square of these two makes a wavelength resolution of 
0.055~$\rm\AA$ (corresponding to $\sim 2.5$ pixels).
Consequently, the resolving power is $R \simeq 10690/0.055 \sim 190,000$.} 
covering 145$''$ (spacial) and 10680--10697~$\rm\AA$ (wavelength) on the 
CCD detector (800 pixels in the dispersion direction and 600 pixels 
vertical to the dispersion direction). 
The data reduction was done by following the standard procedures
(dark subtraction, flat-fielding, spectrum extraction, wavelength
calibration, continuum normalization), where the final 1D spectrum 
was extracted  by integrating over 166 pixels ($= 40''$; i.e., 
$\pm 83$ pixels centered on the target point) in the direction of 
the slit. Although some interference fringes could not be completely 
removed, we could attain S/N ratios on the order of $\sim$100--150, 
which are sufficient for measuring equivalent-widths.

The resulting spectra are displayed in figure 2 (disk center spectrum) 
and figure 3 (12 spectra at different $\mu$\footnote{As usual,
$\mu$ is the direction cosine defined as $\cos \theta$, where $\theta$
is the angle between the line of sight and the normal to the surface
at the observing point.}), where the reference 
disk-center spectrum ($R\sim 350,000$) of the same region taken from
the published solar FTS spectrum atlas\footnote{This spectrum atlas, 
which is based on disk-center observations of the Sun by using the Fourier 
Transform Spectrometer at the McMath telescope of the National Solar 
Observatories at Kitt Peak, is available at 
$\langle$ftp.hs.uni-hamburg.de/pub/outgoing/FTS-Atlas$\rangle$.} 
(Neckel 1994, 1999) is also shown for comparison.
As can be recognized from figure 2, the lines in our disk-center spectrum 
(red symbols) are generally shallower than the reference FTS spectrum
(blue lines). This indicates that our spectra are influenced by 
appreciable scattered light, which must be corrected as will be 
described in subsection 2.3. 

\subsection{Evaluation of Equivalent Widths}

As to evaluating the equivalent widths of three C~{\sc i} lines
at 10683.08, 10685.34, and 10691.24~$\rm\AA$, attention should be paid 
to realizing sufficiently high internal accuracies between the spectra 
of different $\mu$ values.
Given that some lines are partially blended with each other 
(e.g., Si~{\sc i} 10689.72 and C~{\sc i} 10691.12)
and these C~{\sc i} lines show appreciable damping wings, we considered 
that the conventional approach (e.g., direct measurement or Gaussian 
fitting) is not appropriate in the present case.

Therefore, we adopted an alternative approach applied by TTH13 (cf. 
subsections 3.2 and 3.3 therein) based on a spectrum-synthesis modeling.
That is, we first fitted the theoretical synthetic spectrum 
[$I^{\rm th}(\lambda, \mu)$; computed for a given model 
atmosphere] with the observed spectrum [$I^{\rm obs}(\lambda, \mu)$] 
by finding the most optimal solutions of $\log\epsilon$(C) and 
$\log\epsilon$(Si) (C and Si abundances), where the automatic fitting 
algorithm (Takeda 1995a) was applied for this purpose. 
The finally accomplished fit for each spectrum is shown in figure 3.
Then, the equivalent widths ($W_{\lambda}$) of three C~{\sc i} lines 
were {\it inversely} computed from such established solution of 
$\log\epsilon$(C) by using the same model atmosphere.\footnote{
For example, in the case of the C~{\sc i} 10683.08 line, the equivalent 
width ($W'_{10683}$) was computed by using Kurucz's (1993) WIDTH9 
program (curve-of-growth mode) from (1) the solution of $\log\epsilon$(C) 
(resulting from the spectrum fitting at 10680--10695.5~$\rm\AA$;
where the best-fit over the whole region is accomplished by adjusting 
single carbon abundance),  
(2) atomic parameters ($\chi_{\rm low}$ = 7.483~eV and 
$\log gf = +0.076$ for this line; cf. table 2 in TTH13), and
(3) the same $\xi$ and the model atmosphere adopted in the
spectrum-synthesis analysis.
}

\subsection{Scattered-Light Correction}

The apparent $W_{\lambda}$ values derived in the previous subsection 
need to be corrected for scattered light. In order to tackle this problem,
we postulate two assumptions in analogy with the procedure 
adopted by Allende Prieto et al. (2004) and Pereira et al. (2009a):\\
--- (i) The scattered light occurs inside the spectrograph, which 
is wavelength-independent (i.e., not dispersed into spectrum) and 
proportional to the incident radiation (i.e., continuum intensity) 
coming through the slit.\\
--- (ii) The reference disk-center spectrum of solar FTS spectrum atlas
(Neckel 1994, 1999) is not affected by any scattered light.\\

From assumption (i), we can write the scattered light $I_{\rm scat}$
as a fraction $\alpha$ of the continuum intensity $I_{\rm cont}$ 
($I_{\rm scat} = \alpha I_{\rm cont}$).
Then, the ``true'' line depth $R(\lambda)$ and the ``apparent'' line 
depth $R'(\lambda)$ are expressed as follows:
\begin{equation}
R(\lambda) \equiv 1 - \frac{I(\lambda)}{I_{\rm cont}},
\end{equation}
\begin{equation}
R'(\lambda) \equiv 1 - \frac{I(\lambda)+I_{\rm scat}}{I_{\rm cont}+I_{\rm scat}}
      = 1 - \frac{I(\lambda)+\alpha I_{\rm cont}}{I_{\rm cont}+ \alpha I_{\rm cont}}
      = \frac{1}{1+\alpha} \left( 1 - \frac{I(\lambda)}{I_{\rm cont}} \right) .
\end{equation}
Since the true and apparent equivalent widths are defined as 
$W_{\lambda} \equiv \int{R(\lambda)}d\lambda$
and $W'_{\lambda} \equiv \int{R'(\lambda)}d\lambda$, respectively,
equations (1) and (2) lead to the following relation:
\begin{equation}
W'_{\lambda} = W_{\lambda}/(1 + \alpha).
\end{equation}
This means that the correction of $W'_{\lambda}$ is nothing but
a simple multiplication by a constant factor of $1+\alpha$, which equally 
applies to any spectrum of different $\mu$ (note that $\alpha$ is 
a spectrograph-specific quantity at a given wavelength range).

Now, regarding the apparent equivalent widths ($W'_{\lambda}$) of 
C~{\sc i} 10683, 10685, and 10691 lines measured by the method mentioned
in subsection 2.2, we have (241.1, 178.7, and 304.8~m$\rm\AA$)
for the reference disk-center FTS spectrum and 
(196.7, 144.8, and 247.7~m$\rm\AA$) for our DST spectrum at $\mu = 1$.
Taking the former values as ``true'' equivalent widths according
to assumption (ii), we obtain the $W_{\lambda}/W'_{\lambda}$ ratios
($=1+\alpha$) for these three lines as 1.225, 1.234, and 1.231, 
which yield $\alpha = 0.23$ as an average. 

This value was also checked from the viewpoint of spectrum matching. 
Following Allende Prieto et al. (2004) we simulated the line-weakening
effect by adding constant scattered light (fraction $\alpha$ of the 
continuum intensity) to the reference FTS spectrum, which was further
broadened corresponding to the difference of spectral resolution.\footnote{
Since the spectrum resolution of the FTS spectrum is $R_{\rm FTS} \simeq 350,000$ 
and that of our DST spectrum is $R_{\rm DST} \simeq 190,000$, we convolved
$I(\lambda) + \alpha I_{\rm cont}$ with a Gaussian function having a FWHM of
$3\times 10^{5}/R'$~km~s$^{-1}$, where 
$R' (\simeq 230,000)$ was derived from the equation 
$1/R'^{2} = 1/R^{2}_{\rm DST} - 1/R^{2}_{\rm FTS}$.}
We then found that $\alpha \sim 0.25$ (i.e., almost the same as 0.23
derived from $W_{\lambda}$) gives the best fit with our 
disk-center spectrum, as illustrated in figure 4.

Consequently, we adopted the following formula to obtain 
the corrected equivalent widths
\begin{equation}
W_{\lambda}(\mu) \; = \; 1.23 \; W'_{\lambda}(\mu)
\end{equation}
for each of the C~{\sc i} lines. The finally resulting (corrected)
$W_{\lambda}(\mu)$ values are summarized in table~1.\footnote{
For reference, Baschek and Holweger (1967) derived 189~m$\rm\AA$ and
281~m$\rm\AA$ for the disk-center equivalent widths of C~{\sc i} 10685
and 10691 lines, respectively. These are reasonably compared with 
our reference values of 178~m$\rm\AA$ and 305~m$\rm\AA$.}
Yet, we should keep in mind that derivation of these values is based on 
assumptions (i) and (ii), upon which their validity depends.
Nevertheless, as discussed in appendix 1, we will see that our adopted 
$W_{\lambda}$ values of O~{\sc i} 7771--5 triplet lines, for which
the effect of scattered light has been corrected in a similar 
manner ($\alpha = 0.09$ in this case), are consistent with various 
published results (cf. figure 12). This fact may lend support 
for our treatment.

%Sect. 3
\section{Computation of Line Strengths}

\subsection{Model Atmospheres and Microturbulence}

In the calculation of theoretical equivalent widths (at 
various $\mu$ values on the solar disk) to be compared 
with the observational data in table 1, we closely followed the
work of Takeda (1995b; hereinafter referred to as TAK95b) and 
tested two solar atmospheric models ($T_{\rm eff} = 5780$~K, 
$\log g = 4.44$, [Fe/H] = 0.0) having different temperature profiles 
only at the upper layer of $\tau_{5000} \ltsim 10^{-4}$.
Model C has a chromospheric temperature structure of Maltby et al.'s 
(1986) semi-empirical photospheric reference model, while Model E is 
essentially equivalent to Kurucz's (1979) ATLAS6 solar model (without 
any temperature rise). 
The pressure/density structures of these models were obtained 
by integrating the equation of hydrostatic equilibrium. 
See subsection 2.1 of TAK95b (and appendix 2 of this paper) 
for more details about how Models C and E were constructed. 

Regarding the microturbulence ($\xi$), we examined two different 
models: (a) the depth-dependent microturbulent velocity field 
[$\xi_{\rm M}(\tau)$; conspicuously increasing with height at
$\tau_{5000} \ltsim 10^{-4}$] of the reference solar atmosphere
empirically derived by Maltby et al. (1986; cf. their Table 11),
and (b) a depth-dependent microturbulence of 1~km~s$^{-1}$
often adopted in spectroscopic studies of solar spectra (see, e.g.,
subsection 3.2 in Takeda 1994).
The structures of $T(\tau)$, $P(\tau)$, and $n_{\rm e}(\tau)$
for both Models C and E, along with two microturbulence models
($\xi_{\rm M}(\tau)$ and $\xi =$~1~km~s$^{-1}$), are depicted 
in figure 5.
 
\subsection{Non-LTE Calculations}

The procedures of our non-LTE calculations are the same
as adopted in TTH13. As mentioned in section 1,
our main purpose is to examine whether the H~{\sc i} 
collision rates computed with the classical formula are
reasonable or they need to be significantly revised. 
Therefore, as in TTH13, we write the total collisional rates 
($C_{\rm total}$) by introducing a correction factor 
($k_{\rm H}$) as
\begin{equation}
C_{\rm total} = C_{\rm e} + k_{\rm H} C_{\rm H},
\end{equation}
where $C_{\rm e}$ and $C_{\rm H}$ are the collisional rates
due to electrons and neutral hydrogen atoms, respectively, 
computed in the standard recipe (cf. subsection 3.1 of TTH13).

Practically, we tried three $k_{\rm H}$ values (0.1, 1, and 10)
in our calculations, in order to see how the results are 
influenced by changing this parameter.
The way in which the line opacity ($l_{0}$) and the line source
function ($S_{L}$) of the 3s~$^{3}{\rm P}^{\rm o}$ -- 3p $^{3}{\rm D}$
transition (corresponding to C~{\sc i} 10683/10685/10691 lines) 
depend on the choice of $k_{\rm H}$ is displayed in figure 6.
As recognized from this figure, the non-LTE effect is seen in
the increase in the line opacity [$l_{0}$(NLTE)/$l_{0}$(LTE)~$>1$]
as well as in the dilution of the line source function 
($S_{\rm L}/B < 1$) in the line-forming regions; both act in 
the direction of intensifying the lines.  This effect becomes 
progressively more conspicuous with a decrease in $k_{\rm H}$, 
as naturally expected because a decrease in the collisional rates 
tends to enhance the departure from LTE.

%Sect. 4
\section{Discussion}

\subsection{Effect of Atmospheric Structure}

We first discuss the effects caused by different choices of solar 
atmospheric model as well as of microturbulence on the equivalent 
widths of C~{\sc i} lines computed at various $\mu$ values.
Figure 7 shows the non-LTE $W_{\lambda}$ vs. $\mu$ relations
corresponding to $k_{\rm H} = 1$, which were computed 
with a fixed carbon abundance for Model C (red thick lines) 
and Model E (blue thin lines) in combination with 
$\xi = \xi_{\rm M}(\tau)$ (solid lines) and 
$\xi = 1$~km~s$^{-1}$ (dashed lines).
We can recognize the following characteristics from this figure:\\
--- First, the difference between two microturbulence models, 
depth-dependent $\xi_{\rm M}(\tau)$ and constant $\xi$~=~1~km~s$^{-1}$, 
causes only insignificantly small changes in $W_{\lambda}$, 
which is presumably because that these two $\xi$ models do not 
differ much from each other (in the average sense) at the important 
line-forming region ($10^{-3} \ltsim \tau_{5000} \ltsim 1$; cf. figure 5).\\
--- Second, an appreciable difference is seen between the results
of Model C and Model E; that is, $W_{\lambda}$(C) tends to be larger
than $W_{\lambda}$(E), and the difference systematically increases
toward the limb (i.e., with a decrease in $\mu$). This is just the
same tendency found in the case of O~{\sc i} 7771--5 triplet lines
as pointed out by TAK95b (cf. figure 12 therein). Since C~{\sc i} 
10683/10685/10691 lines and O~{\sc i} 7771--5 lines share the 
similar characters to each other (i.e., strong high-excitation 
lines of multiplet 1 originating from a metastable term), this 
trend can be interpreted as due to the same mechanism working 
for those O~{\sc i} lines (i.e., shift of the core forming 
region upward into the higher layer where the line source function 
is more diluted, since the line opacity is enhanced due to the 
temperature rise; cf. subsection 2.4 in TAK95b).

\subsection{Influence of Changing $k_{H}$}

The effect of changing $k_{\rm H}$ is demonstrated in figure 8,
where the $W_{\lambda}$ vs. $\mu$ relations for the representative
C~{\sc i} 10683 line, corresponding to $k_{\rm H}$ = 0.1, 1, 10,
and $\infty$ (LTE), are depicted for two solar model atmospheres;
Model~C with $\xi = \xi_{\rm M}(\tau)$ and Model~E with 
$\xi = 1$~km~s$^{-1}$).
It can be seen from figure 8a (fixed carbon abundance) that 
$W_{\lambda}$ at a given $\mu$ tends to increase with a decrease 
in $k_{\rm H}$, since the non-LTE effect (being enhanced with a 
lowered $k_{\rm H}$) always acts in the direction of strengthening
a line as explained in subsection 3.2.
This effect becomes more appreciable for smaller $\mu$
(toward the limb), because the line-forming region systematically 
moves toward an upper layer with a decrease in $\mu$ where the 
non-LTE effect is stronger.
This figure also exhibits the effect of $T(\tau_{5000})$ structure
mentioned in the previous subsection; i.e., 
the inequality relation of $W_{\lambda}$(C)~$>$~$W_{\lambda}$(E) 
at a given $\mu$ and $k_{\rm H}$.\footnote{Note that this inequality
exceptionally breaks down for the case of LTE, where 
$W_{\lambda}$(C)~$<$~$W_{\lambda}$(E) holds especially
near to the limb, since the chromospheric temperature rise in  
Model~C causes a core emission in the line profile when computed
with LTE, which eventually makes the equivalent width smaller.}

In comparing the results calculated with various conditions with 
observations, it is necessary to appropriately adjust the carbon 
abundance at the disk center so that the condition 
$W_{\lambda}^{\rm cal}(\mu = 1)$ = 
$W_{\lambda}^{\rm obs}(\mu = 1)$ holds for each case.
Such normalized $W_{\lambda}$ vs. $\mu$ relations after this 
matching has been applied are presented in figure 8b, where
we can see that the gradient $|dW_{\lambda}/d\mu|$ is lessened
with a lowering of $k_{\rm H}$, reflecting the that non-LTE
intensification of $W_{\lambda}$ is enhanced with decreasing $\mu$ 
toward the limb. This characteristics may be used to give a
constraint on $k_{\rm H}$ in comparison with the observed 
$\mu$-dependence of $W_{\lambda}$.

\subsection{Comparison with Observations and Conclusion}

We are now ready to get empirical information on $k_{\rm H}$
by comparing the theoretical and observed $W_{\lambda}$ vs. $\mu$ 
relations: the primary aim of this study. 
Here, we exclusively adopt Model~C with $\xi = \xi_{\rm M}(\tau)$ 
as the standard solar atmospheric model since it is considered 
to be comparatively more realistic, while keeping in mind that 
an application of Model~E causes a marginal increase in the
gradient $|dW_{\lambda}/d\mu|$ (i.e., equivalent to an increase in 
$k_{\rm H}$) as compared to the case of Model~C (cf. figure 8b). 

Our results are shown in figure 9, where the observed
center-to-limb variations of $W_{\lambda}$ for C~{\sc i}
10683/10685/10691 lines  are overplotted with 
the theoretical relations corresponding to $k_{\rm H}$ = 
0.1, 1, 10, and $\infty$ (LTE). 
This figure indicates that the observed data are almost
consistent with the relations for $k_{\rm H} =1$ in the 
average/global sense (especially based on the matching at 
$0.6 \gtsim \mu \gtsim 0.3$), though it is difficult to establish 
an exact solution because of the rather large scatter comparable
to the difference caused by changing $k_{\rm H}$.
We may thus state that non-LTE calculations 
using the H~{\sc i} collision rates evaluated with the formula 
based on the classical cross section can reasonably reproduce 
the observed $W_{\lambda}$ vs. $\mu$ relations of these C~{\sc i} 
lines, which means that the necessity of significantly revising 
these classical $C_{\rm H}$ rates (e.g., considerable
reduction, as occasionally argued) is unlikely.
This is the conclusion of this investigation.

\subsection{Implication and Problems}

This consequence has a significant implication on carbon abundance
determinations in very metal-poor stars. TTH13 found that
C abundances derived from CH lines tend to be smaller than
those derived from C~{\sc i} 10683/10685/10691 lines by several 
tenths dex for stars at [Fe/H]~$\ltsim -2$. 
One possible scenario that might explain this discrepancy was to 
considerably reduce $k_{\rm H}$ (from the value of unity assumed in TTH13),
since it would further lower the C~{\sc i} abundance (due to an
enhanced non-LTE effect) in the direction of mitigating the
discordance. Now that this possibility has been ruled out, we feel 
that this disagreement between C~{\sc i} and CH abundances may be 
rather attributed to an underestimation of the latter, since 
molecular lines are considerably sensitive to temperature 
structures of upper layers and apt to be unreliable.

Here, a remark of caution may be relevant. 
It may still be rather premature to regard this 
consequence as conclusive. We should realize that our result is 
based on a classical Kurucz's (1979) solar atmospheric model  
(though the chromospheric effect was taken into account 
by referring to Maltby et al.'s model), while other published
photospheric models have slightly different structures
from that we adopted (cf. appendix 2).
In apppendix 1, the $k_{\rm H}$ value relevant for O~{\sc i} 
7771--5 lines is discussed from the $\mu$-dependence of 
$W_{\lambda}$, where our theoretical calculations are done
with the same solar model atmosphere.
As we will see, we obtained a result discrepant from that  
obtained recently by Pereira et al. (2009b) based on their 
state-of-art 3D dynamical solar model atmosphere.
It is interesting to see, therefore, how the present result
may be influenced by applying this kind of new approach. 
Yet, since the use of such a realistic 3D model tends to
raise the theoretical $W_{\lambda}$ of these C~{\sc i} lines 
(i.e., equivalent to decreasing $k_{\rm H}$) according to 
Asplund et al. (2005), which is also expected from the case of 
O~{\sc i} 7771--5 lines having similar excitation potentials 
(cf. figure 5 in Pereira et al. 2009b), its application 
would act in the direction of increasing the empirical solution 
for $k_{\rm H}$. This makes the possibility of reducing 
$k_{\rm H}$ even more unlikely.

Finally, we should recall that our adopted equivalent widths were 
derived by applying appreciable scattered light corrections. 
Thus, follow-up independent observations of center-to-limb variations of 
these C~{\sc i} 1.07~$\mu$m lines (preferably with an instrument such 
as FTS) would be desired to check our correction procedure.

%Appendix 1 & 2
\appendix

\section{Center-to-Limb Variation of O~I 7771$-$5 $\rm\AA$ Lines}

On the same day of 2013 July 26 when our solar spectroscopic 
observations of C~{\sc i} 10683/10685/10691 lines were carried out, 
we also observed the O~{\sc i} 7771--5 triplet lines, which share 
similar characters to those of C~{\sc i} lines (e.g, strong 
lines of multiplet 1 originating from high-excited metastable term). 
This is because we want to check our observation and data reduction 
process (especially in terms of the procedure for scattered-light 
correction), since empirical $W_{\lambda}$ vs. $\mu$ relations for 
these O~{\sc i} lines have already been published by 
several investigators and can be compared with ours.

We found that the lines in our disk-center spectrum are generally 
weaker than the reference FTS spectrum also for this case of 
O~{\sc i} lines as shown in figure 10, though the extent of weakening
is comparatively milder than the case of C~{\sc i} lines, which 
indicates that the effect of scattered light in our instrument 
becomes more appreciable with increasing wavelength.
Since the apparent equivalent widths ($W'_{\lambda}$) of these O~{\sc i} 
7771, 7774, and 7775 lines evaluated in the same way using synthetic
spectrum fitting (cf. figure 11) as described in subsection 2.2 turned 
out to be (83.6, 72.1, and 57.1~m$\rm\AA$) for the reference 
disk-center FTS spectrum and (77.1, 66.2, and 52.0~m$\rm\AA$) for 
our DST spectrum at $\mu = 1$, the resulting $W_{\lambda}/W'_{\lambda}$ 
ratios ($=1+\alpha$) are 1.084, 1.089, and 1.098, respectively, 
which yield $\alpha = 0.09$ as an average. 
Accordingly, we multiplied the apparent $W'_{\lambda}$ values 
by a correction factor of 1.09 to obtain the final equivalent widths,
which are summarized in table 2.

These $W_{\lambda}$ results are plotted against $\mu$ in figure 12, 
where the published values taken from various literature are also shown.
We can state by inspecting this figure that the equivalent widths derived
from our observation are mostly in reasonable agreement with 
those published ones, which may justify our measurement 
as well as the correction procedure for stray light.

In analogy with figure 9, theoretical relations for these 
O~{\sc i} 7771--5 lines corresponding to $k_{\rm H}$ = 0.1, 1, 10, 
and $\infty$ (LTE) computed for Model~C with $\xi = \xi_{\rm M}(\tau)$ 
are also depicted in figure 12. This figure indicates that all these 
theoretical $W_{\lambda}$ vs. $\mu$ relations calculated for three
$k_{\rm H}$ values situate below the observed data; i.e.,
we would have to require $k_{\rm H} < 0.1$ in order to bring 
theory and observation into agreement. This is incompatible
with the consequence of TAK95a, where $k_{\rm H} \sim 1$ was suggested
based on the requirement of abundance consistency between O~{\sc i} 
7771--5 lines and other O~{\sc i} lines (see also Takeda \& Honda 
2005; cf. appendix 1 therein).
Actually, TAK95a could not reproduce the observed $W_{\lambda}$ 
vs. $\mu$ relation for the O~{\sc i} 7771 line quantitatively 
well by his non-LTE calculations done with $k_{\rm H} = 1$, 
though he pointed out that inclusion of chromospheric temperature 
rise can somewhat mitigate the discrepancy (cf. figure 12 therein). 
Thus, we must admit that there is still a disagreement in the value
of $k_{\rm H}$ for the O~{\sc i} 7771--5 lines between these
two approaches, as far as our adopted model atmosphere is 
concerned.\footnote{
However, based on Kurucz's (1993) ``non-overshooting'' 
ATLAS9 model (which should not be so different from our adopted
model), Allende Prieto et al. (2004) reported that $k_{\rm H} = 1$
(rather than $k_{\rm H} = 0$) better reproduces the observed 
$W_{\lambda}$ vs. $\mu$ relation of the O~{\sc i} 7771--5 lines
(cf. their figure 6), which contradicts our result. Though the 
reason is not clear, some difference might exist between our and 
their non-LTE calculation procedures.
}

Nevertheless, this problem may be resolved by applying different 
models, since the result seems sensitive to adopted atmospheric 
structures. For example, Pereira et al. (2009b) showed that 
the choice of $k_{\rm H} \simeq 1$ can accomplish
the best agreement between theoretical and observed center-to-limb
variations of O~{\sc i} 7771--5 lines when their sophisticated
3D hydrodynamical model was used,\footnote{Interestingly, the 
application of Holweger and M\"{u}ller's (1974) empirical solar 
model seems to yield a result similar to the case of the recent 
3D model, as seen from the comparison of Pereira et al.'s (2009b)
figures 4 (upper-middle panel) and 5 (upper-right panel).}
while no solution could be obtained ($k_{\rm H} < 0.01$, just like 
our case) when Gustafsson et al.'s (2008) 1D MARCS model was used. 
Figure 5 of Pereira et al. (2009b) illustrates this situation, 
where we can confirm that our figure 12b is quite similar 
to the upper-left panel (corresponding to MARCS model) of their 
figure 5. Therefore, the circumstance is rather complicated
depending on the adopted model atmosphere, which led us
to remark in subsection 4.4 that applying the new 3D model 
atmosphere would be desirable to check our conclusion.

\section{Comparison of Temperature Structures of Various Solar 
Photospheric Models}

As in TAK95b, the solar model atmospheres adopted in this study 
(Model~E and Model~C) are based on Kurucz's (1979) theoretical model 
computed by his ATLAS6 program, where the convection was taken into 
account by the local mixing-length theory with $l = 2 H_{\rm p}$
($H_{\rm p}$: pressure scale height). That is, this ATLAS6 solar 
model atmosphere (defined at $-4 \ltsim \log \tau_{5000}$)
was simply extrapolated at $-7 \ltsim \log \tau_{5000} \ltsim -4$ 
for Model~E, while the chromospheric $T(\tau_{5000})$ structure of
Maltby et al.'s (1986) photospheric reference model was assumed at
$-7 \ltsim \log \tau_{5000} \ltsim -4$ (and the hydrostatic equation 
was integrated to obtain the pressure structure) for Model~C.

As shown in this study as well as in TAK95b, the difference in the 
structure of upper optically-thin layer (i.e., between Model~E and Model~C) 
does not necessarily have so large impact on the resulting $W$ vs. 
$\mu$ relations of C~{\sc i} 10683/10685/10691 and 
O~{\sc i} 7771/7774/7775 lines in the quantitative sense, though 
the changes are surely appreciable.  

On the other hand, the temperature structure of deeper photospheric layers
is considered to influence more significantly on the formation of
these C~{\sc i} and O~{\sc i} lines of high-excitation. Actually,
as seen from the distribution of contribution function of these lines,  
most contribution comes from $-1 \ltsim \log \tau_{5000} \ltsim +0.5$
(cf. figure 7 of Takeda 1994 for the case of disk center, though 
the formation layer is systematically shifted upward at the limb).
Accordingly, it is worthwhile to examine how the photospheric 
$T(\tau_{5000})$ structure of Kurucz's (1979) ATLAS6 model, on which 
our Model~E and Model~C are based, are compared with those of other 
representative solar model atmospheres published so far.   

Such comparisons of temperature structures of different solar photospheric 
models are displayed in figure 13, where Holweger and M\"{u}ller's (1974) 
empirical model, Kurucz's (1979) theoretical ATLAS6 model, Maltby et al's 
(1986) semi-empirical photospheric reference model, Gustafsson et al.'s (2008) 
theoretical MARCS model, Kurucz's (1993) theoretical ATLAS9 model 
(with convective overshooting), and the new 3D dynamical model used 
by Pereira et al. (2009b) (averaged $\langle T(\tau_{5000}) \rangle$) 
are compared with each other.

An inspection of this figure suggests that the $T(\tau_{5000})$ structures
of these models do not differ so much (differences are on the order of 
$\sim 100$~K) except for the convection-dominated deeper layer 
($\log \tau_{5000} \gtsim +0.5$).
Yet, it should be realized that a temperature change of $\sim$~100~K
suffices to cause an appreciable effect on the strength of these high-excitation 
C~{\sc i} and O~{\sc i} lines in question, as can be recognized from
Pereira et al's (2009b) results for oxygen lines. In this sense,
which solar photospheric model to choose is an important factor 
to be kept in mind when interpreting or discussing the matching of 
theoretical/empirical $W$ vs. $\mu$ relation.

%Table 1
\clearpage
\setcounter{table}{0}
\begin{table}[h]
%\scriptsize
\small
\caption{Equivalent widths of C~{\sc i} 10683.08, 10685.34, 
and 10691.24 lines at various $\mu$ points on the solar disk.}
\begin{center}
\begin{tabular}
{ccccc}\hline \hline
$d$ & $\mu$ & $W_{10683}$ & $W_{10685}$ & $W_{10691}$ \\
($R_{0}$) & & (m$\rm\AA$) & (m$\rm\AA$) & (m$\rm\AA$) \\
\hline
  0.00 & 1.000 &  242 & 178 & 305 \\
  0.45 & 0.893 &  220 & 161 & 277 \\
  0.50 & 0.866 &  224 & 164 & 282 \\
  0.55 & 0.835 &  222 & 163 & 280 \\
  0.60 & 0.800 &  219 & 160 & 275 \\
  0.65 & 0.760 &  214 & 157 & 270 \\
  0.70 & 0.714 &  210 & 154 & 265 \\
  0.75 & 0.661 &  213 & 156 & 269 \\
  0.80 & 0.600 &  198 & 144 & 250 \\
  0.85 & 0.527 &  191 & 139 & 241 \\
  0.90 & 0.436 &  194 & 141 & 245 \\
  0.95 & 0.312 &  174 & 125 & 220 \\
\hline
\end{tabular}
\end{center}
\scriptsize
Note. \\
Given here are the finally adopted values after being corrected for 
the effect of scattered light by multiplying the correction
factor of 1.23. (The directly measured raw equivalent widths
are obtained by dividing these values by the same factor.)
The apparent distance $d$ on the solar disk between the observed point 
and the disk center (in unit of $R_{0}$, the apparent radius
of the disk) is connected with 
$\mu$ ($\equiv \cos\theta$; direction cosine) by the relation 
$\mu = \sqrt{1-d^{2}}$.
\end{table}

%Table 2
\clearpage
\setcounter{table}{1}
\begin{table}[h]
%\scriptsize
\small
\caption{Equivalent widths of O~{\sc i} 7771.94, 7774.17, 
and 7775.39 lines at various $\mu$ points on the solar disk.}
\begin{center}
\begin{tabular}
{ccccc}\hline \hline
$d$ & $\mu$ & $W_{7771}$ & $W_{7774}$ & $W_{7775}$ \\
($R_{0}$) & & (m$\rm\AA$) & (m$\rm\AA$) & (m$\rm\AA$) \\
\hline
  0.00 & 1.000 &  84 &  72 &  57 \\
  0.45 & 0.893 &  81 &  70 &  55 \\
  0.50 & 0.866 &  79 &  68 &  53 \\
  0.55 & 0.835 &  78 &  67 &  53 \\
  0.60 & 0.800 &  77 &  66 &  52 \\
  0.65 & 0.760 &  77 &  66 &  52 \\
  0.70 & 0.714 &  75 &  64 &  50 \\
  0.75 & 0.661 &  72 &  62 &  48 \\
  0.80 & 0.600 &  72 &  62 &  48 \\
  0.85 & 0.527 &  67 &  57 &  44 \\
  0.90 & 0.436 &  63 &  54 &  42 \\
  0.95 & 0.312 &  61 &  51 &  40 \\
\hline
\end{tabular}
\end{center}
\scriptsize
Note. \\
Given here are the finally adopted values after being corrected for 
the effect of scattered light by multiplying the correction
factor of 1.09. (The directly measured raw equivalent widths
are obtained by dividing these values by the same factor.)
\end{table}

\clearpage

%Figure 1
\begin{figure}
  \begin{center}
    \FigureFile(100mm,100mm){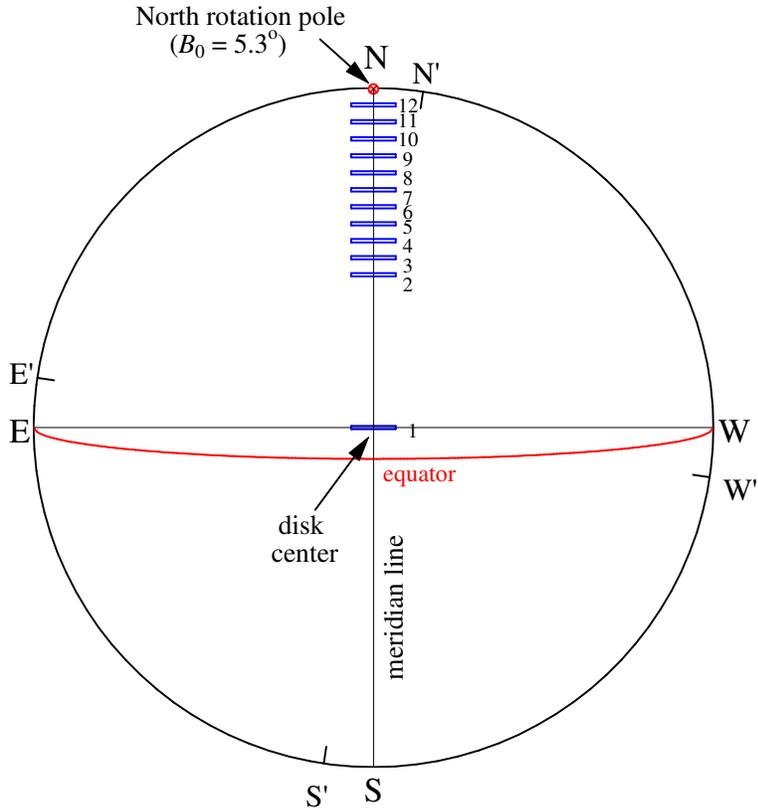}
    %%% \FigureFile(width,height){filename}
  \end{center}
\caption{
Observed 12 points on the meridian line of the solar disk 
(1 point at the disk center and 11 points from 0.45~$R_{0}$ to 
0.95~$R_{0}$ with a step of 0.05~$R_{0}$, where $R_{0}$ is the
apparent radius of the solar disk), at which the slit was 
aligned in the E--W direction. While N, S, E, and W are the directions
in reference to the Sun (based on solar rotation), those in the 
equatorial coordinate system on the celestial sphere (defined by 
the rotation of Earth) are also denoted as N$'$, S$'$, E$'$, and W$'$.
}
\end{figure}

%Figure 2
\begin{figure}
  \begin{center}
    \FigureFile(120mm,100mm){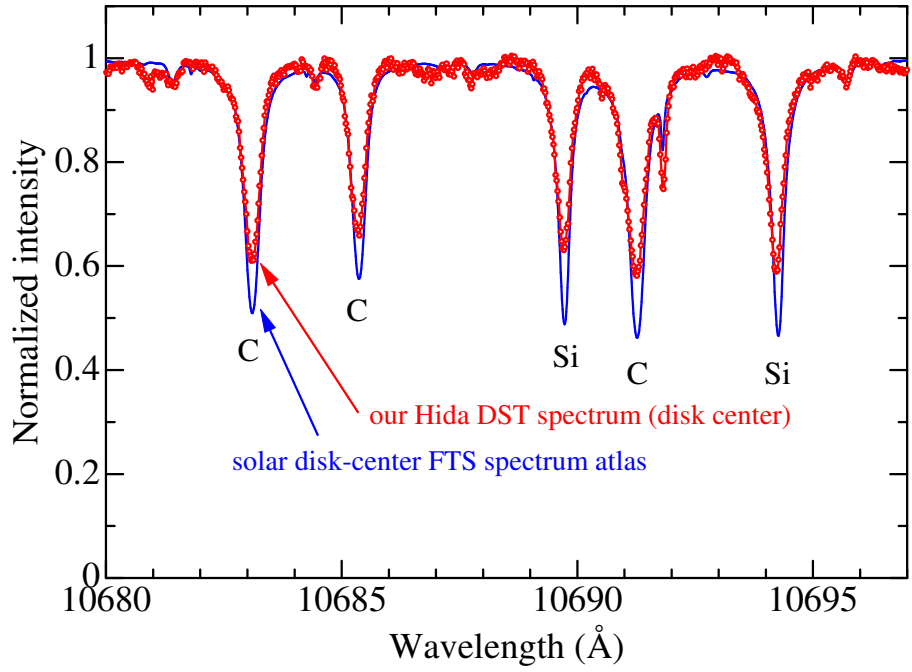}
    %%% \FigureFile(width,height){filename}
  \end{center}
\caption{
Comparison of our 10680--10697~$\rm\AA$ region spectrum (comprising
C~{\sc i} and Si~{\sc i} lines) at the disk center 
with that of solar disk-center FTS spectrum atlas (Neckel 1994, 1999),
where the former is displayed in (red) open circles connected by lines
and the latter is in (blue) solid lines.
}
\end{figure}

%Figure 3
\begin{figure}
  \begin{center}
    \FigureFile(120mm,160mm){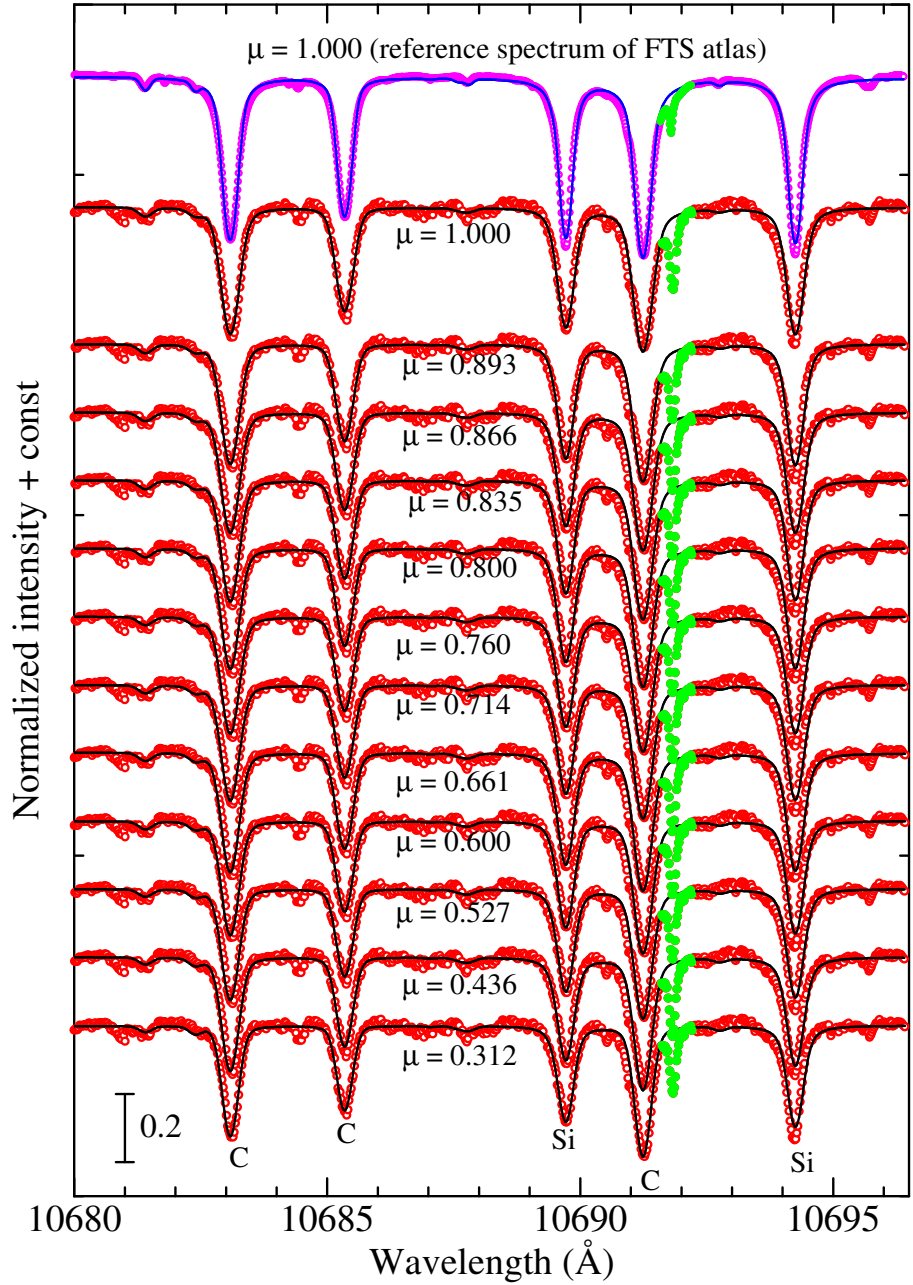}
    %%% \FigureFile(width,height){filename}
  \end{center}
\caption{
Synthetic spectral fitting for each spectrum of different $\mu$ 
accomplished by finding the best-match abundance solution of C 
(and Si), from which the equivalent widths of C~{\sc i} 
10683/19685/10691 lines were inversely computed.
The theoretical spectra are shown by solid lines while 
the observed data are plotted by circles (where those of telluric 
origin and masked in the fitting are highlighted in green).
Following the reference FTS spectrum displayed
at the top, our DST spectra are arranged in 
the descending order of $\mu$ with appropriate offsets relative 
to the adjacent ones.
}
\end{figure}

%Figure 4
\begin{figure}
  \begin{center}
    \FigureFile(120mm,100mm){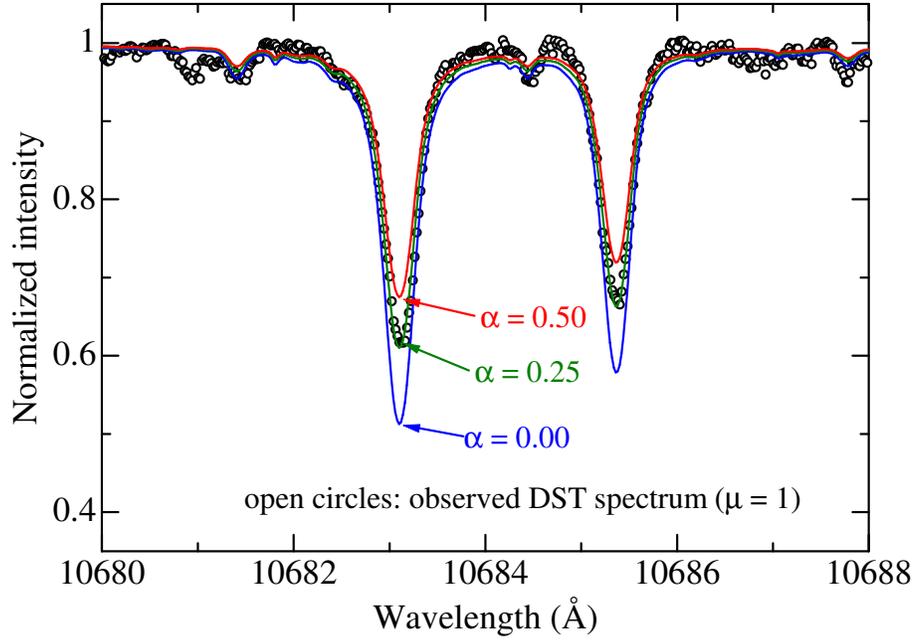}
    %%% \FigureFile(width,height){filename}
  \end{center}
\caption{
Simulation of how the inclusion of scattered light affects the
profiles of C~{\sc i} 10683 and 10685 lines in the disk-center spectrum.
The three solid lines show the simulated spectra computed 
based on the reference solar disk-center FTS spectrum,
to which three values of scattered light (fraction $\alpha$ of
the continuum intensity; $\alpha$ = 0.00, 0.25, and 0.50 are colored 
in blue, green, and red, respectively) were added and an appropriate 
broadening was applied to take into account the difference of spectral 
resolution (see subsection 2.3). Our DST spectrum ($\mu =1$) is also 
shown for comparison (open circles). 
}
\end{figure}

%Figure 5
\begin{figure}
  \begin{center}
    \FigureFile(120mm,160mm){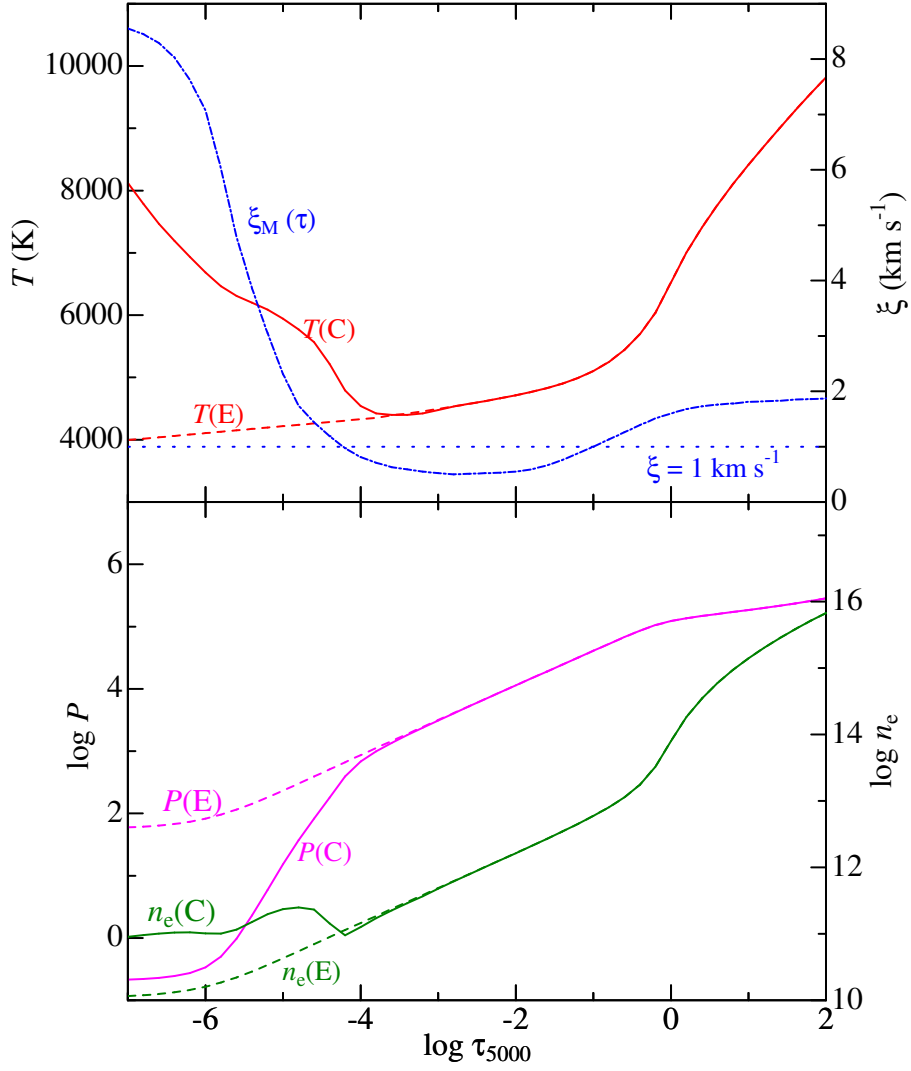}
    %%% \FigureFile(width,height){filename}
  \end{center}
\caption{
Distributions of $T$ (temperature), $P$ (pressure), and $n_{\rm e}$
(electron density) plotted against $\log \tau_{5000}$ for two solar 
atmospheric models, Model C (solid lines) and Model E (dashed lines), 
used in this study (cf. subsection 3.1).
In addition, Maltby et al.'s (1986) depth-dependent microturbulence 
[$\xi_{\rm M}(\tau)$; dash-dotted line] for the reference solar atmosphere
is compared with the occasionally used (depth-independent) value 
of $\xi =$~1~km~s$^{-1}$ (dotted line). 
}
\end{figure}

%Figure 6
\begin{figure}
  \begin{center}
    \FigureFile(120mm,160mm){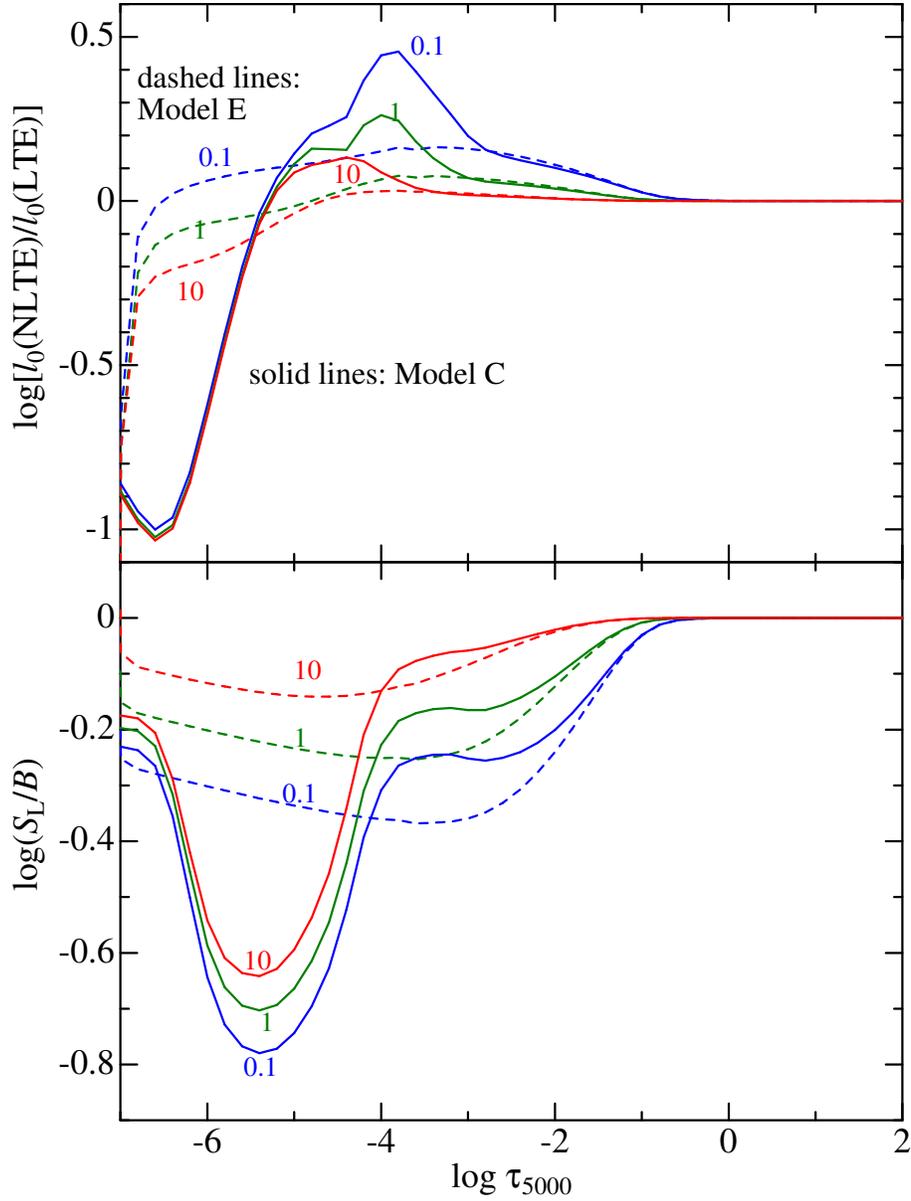}
    %%% \FigureFile(width,height){filename}
  \end{center}
\caption{
Depth-dependence of the non-LTE to LTE opacity ratio (upper panel)
as well as of the line source function ($S_{\rm L}$) in unit of
the local Planck function ($B$) (lower panel), calculated for 
three $k_{\rm H}$ values (0.1, 1, and 10) as indicated in the figure.
The results for Model C are depicted in solid lines, while 
those for Model E are in dashed lines.
} 
\end{figure}

%Figure 7
\begin{figure}
  \begin{center}
    \FigureFile(120mm,80mm){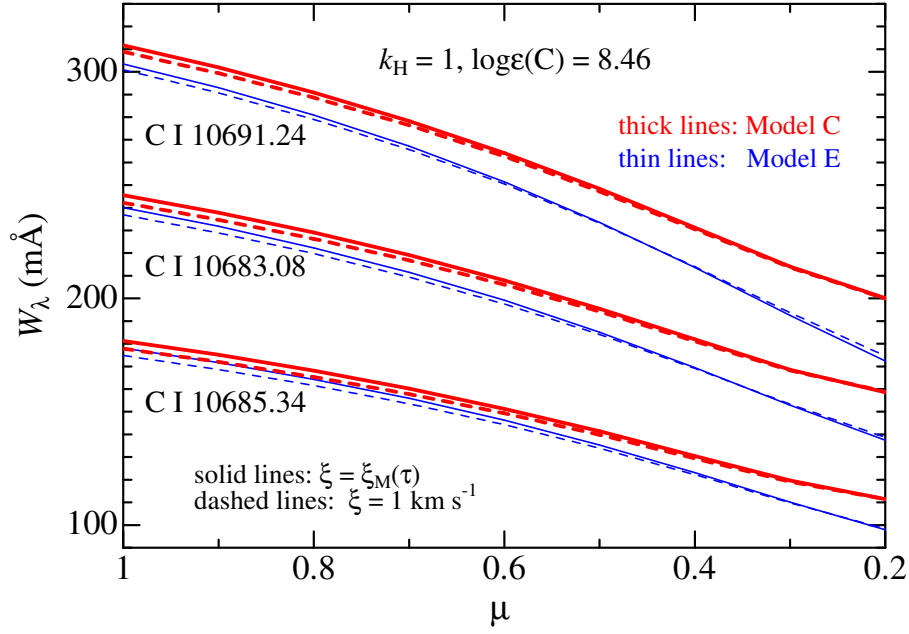}
    %%% \FigureFile(width,height){filename}
  \end{center}
\caption{
Demonstration of how the difference in the model structure or in the 
microturbulence affects on the $W_{\lambda}$ vs. $\mu$ relations 
of C~{\sc i} 10683/10685/10691 lines.
Presented are the non-LTE results corresponding to $k_{\rm H} = 1$,
which were computed with $\log\epsilon$(C) = 8.46 for Model C 
(red thick lines) and Model E (blue thin lines) in combination 
with $\xi = \xi_{\rm M}(\tau)$ (solid lines) and 
$\xi = 1$~km~s$^{-1}$ (dashed lines).
}
\end{figure}

%Figure 8
\begin{figure}
  \begin{center}
    \FigureFile(120mm,160mm){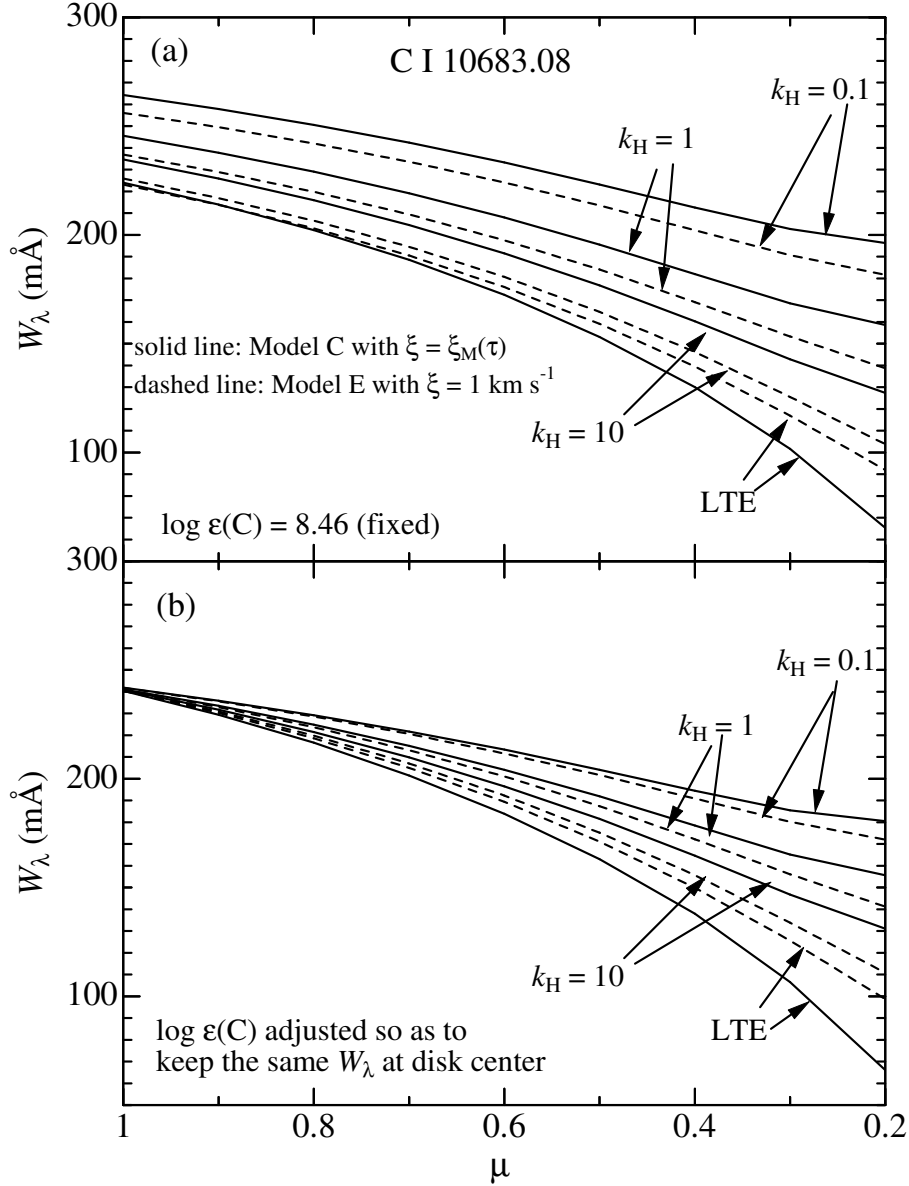}
    %%% \FigureFile(width,height){filename}
  \end{center}
\caption{
Demonstration of how the change in $k_{\rm H}$ affects on the 
$W_{\lambda}$ vs. $\mu$ relation of the representative 
C~{\sc i} 10683.08 line. Computations were done for $k_{\rm H}$ = 
0.1, 1, 10, and $\infty$ (LTE) as indicated in the figure. 
The results for Model C with $\xi = \xi_{\rm M}(\tau)$ are
shown in solid lines, and those for Model E with 
$\xi = 1$~km~s$^{-1}$ are in dashed lines.
The relations derived for a fixed carbon abundance of 
$\log\epsilon$(C) = 8.46 are presented in the upper panel (a), 
while those obtained with appropriately adjusted 
carbon abundances (to keep $W_{\lambda}(\mu = 1)$ at the same 
value of 242~m$\rm\AA$) are given in the lower panel (b).
}
\end{figure}

%Figure 9
\begin{figure}
  \begin{center}
    \FigureFile(120mm,160mm){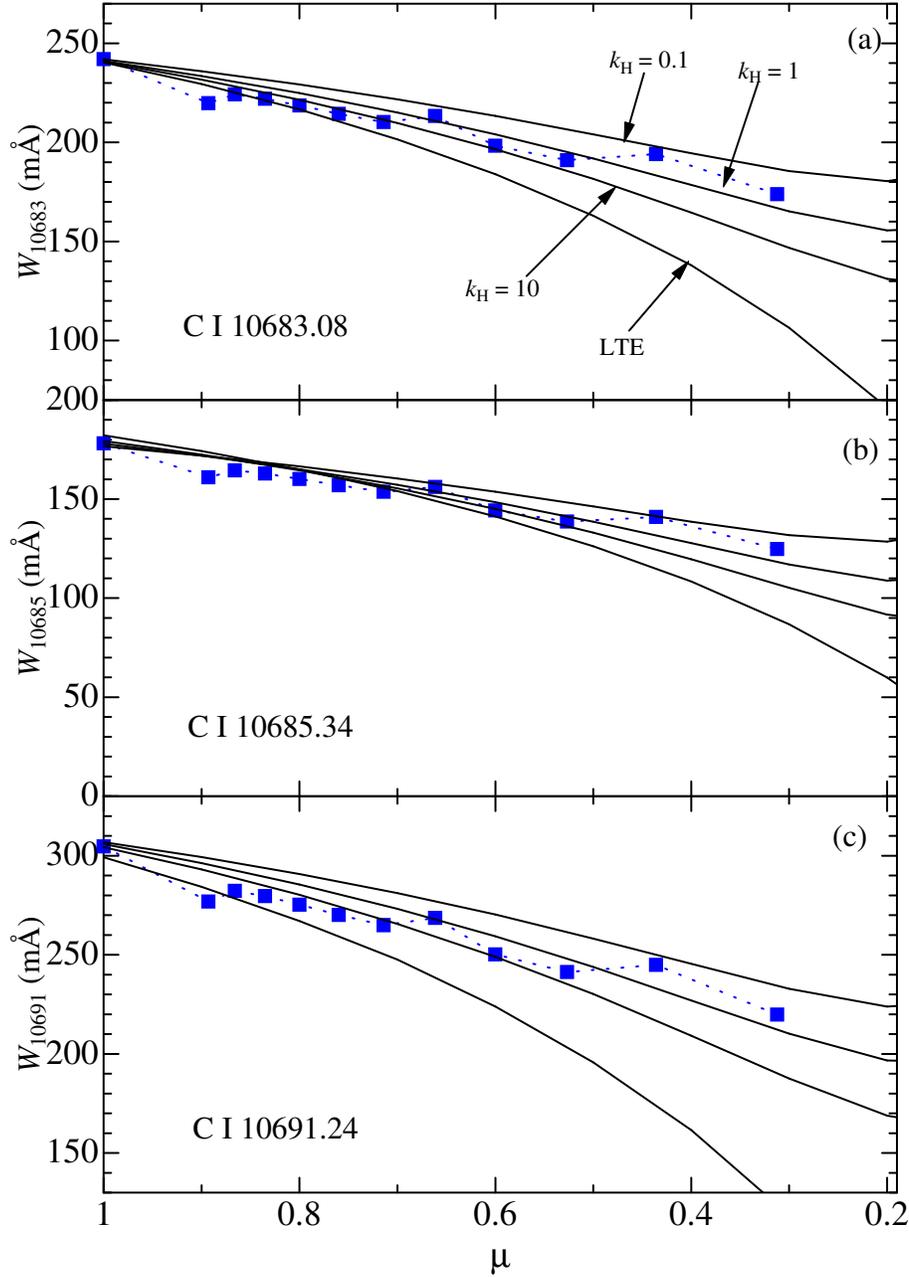}
    %%% \FigureFile(width,height){filename}
  \end{center}
\caption{
Comparison of the theoretical $W_{\lambda}$ vs. $\mu$ relations
(solid lines) computed for Model C with $\xi = \xi_{\rm M}(\tau)$ 
with the observed center-to-limb variation of the equivalent 
widths (filled squares; cf. table 1).
Panels (a), (b), (c) present the results for C~{\sc i} 10683.08,
10685.34, and 10691.24 lines, respectively.
Calculations were done for $k_{\rm H}$ = 0.1, 1, 10, and $\infty$ (LTE) 
(generally, $W_{\lambda}$ decreases with an increase in $k_{\rm H}$ 
at a given $\mu$) as indicated in the figure, and the abundances 
were adequately adjusted each case so as to make the computed
$W_{\lambda}$($\mu =1$) almost consistent with the observed 
disk-center values.
}
\end{figure}

%Figure 10
\begin{figure}
  \begin{center}
    \FigureFile(120mm,100mm){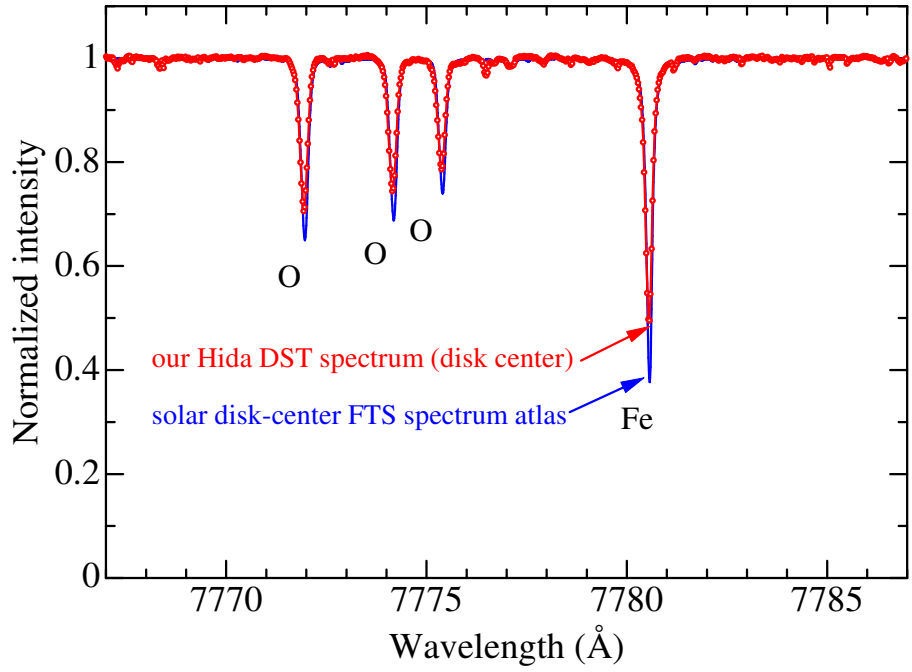}
    %%% \FigureFile(width,height){filename}
  \end{center}
\caption{
Comparison of our 7767--7787~$\rm\AA$ region spectrum (comprising 
O~{\sc i} and Fe~{\sc i} lines) at the disk center with that of 
solar disk-center FTS spectrum atlas. Otherwise, the same as
in figure 2.
}
\end{figure}

%Figure 11
\begin{figure}
  \begin{center}
    \FigureFile(120mm,160mm){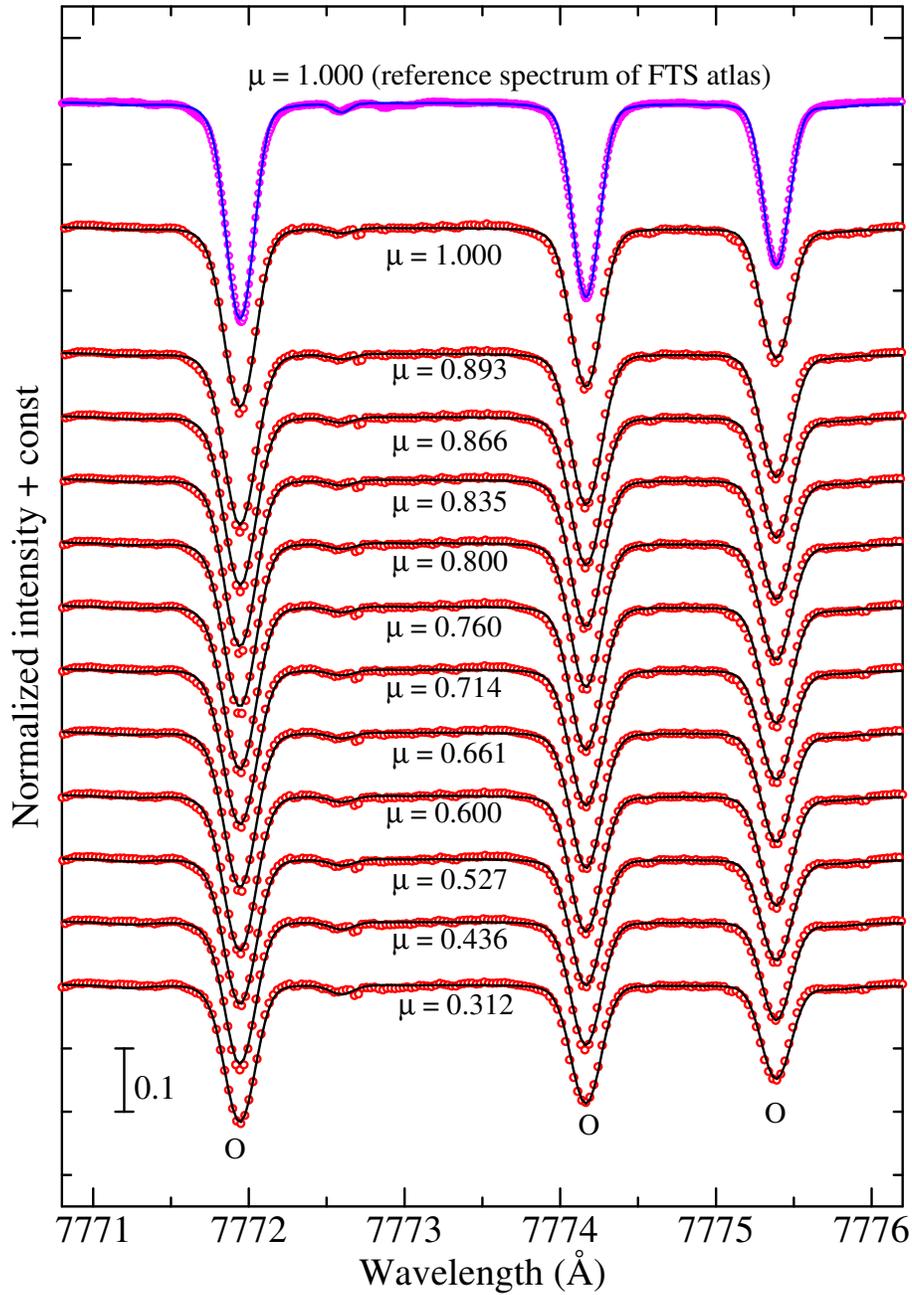}
    %%% \FigureFile(width,height){filename}
  \end{center}
\caption{
Synthetic spectral fitting for each spectrum of different $\mu$ 
accomplished by finding the best-match abundance solution of O, 
from which the equivalent widths of O~{\sc i} 7771/7774/7775 lines were 
inversely computed. Otherwise, the same as in figure 3.
}
\end{figure}

%Figure 12
\begin{figure}
  \begin{center}
    \FigureFile(120mm,160mm){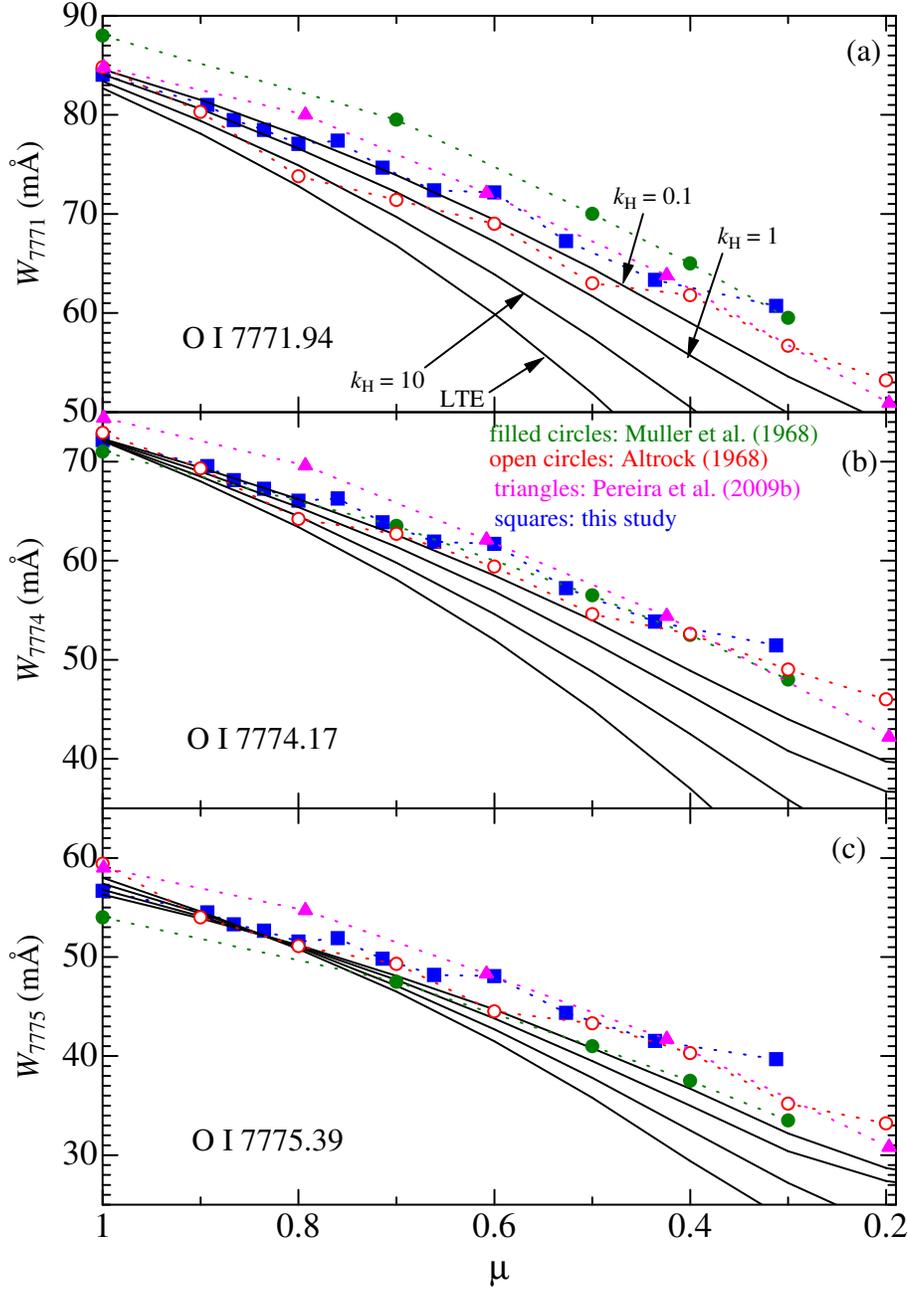}
    %%% \FigureFile(width,height){filename}
  \end{center}
\caption{
Comparison of the theoretical $W_{\lambda}$ vs. $\mu$ relations
(solid lines) computed for Model C with $\xi = \xi_{\rm M}(\tau)$ 
with the observed center-to-limb variation of the equivalent 
widths collected from various literature:
Filled circles $\cdots$ M\"{u}ller et al. (1968),
open circles $\cdots$ Altrock (1968), 
filled triangles $\cdots$ Pereira et al. (2009b), and
filled squares $\cdots$ our measurements.
Panels (a), (b), (c) present the results for O~{\sc i} 7771.94,
7774.17, and 7775.39 lines, respectively.
Otherwise, the same as in figure 9.
}
\end{figure}

%Figure 13
\begin{figure}
  \begin{center}
    \FigureFile(120mm,160mm){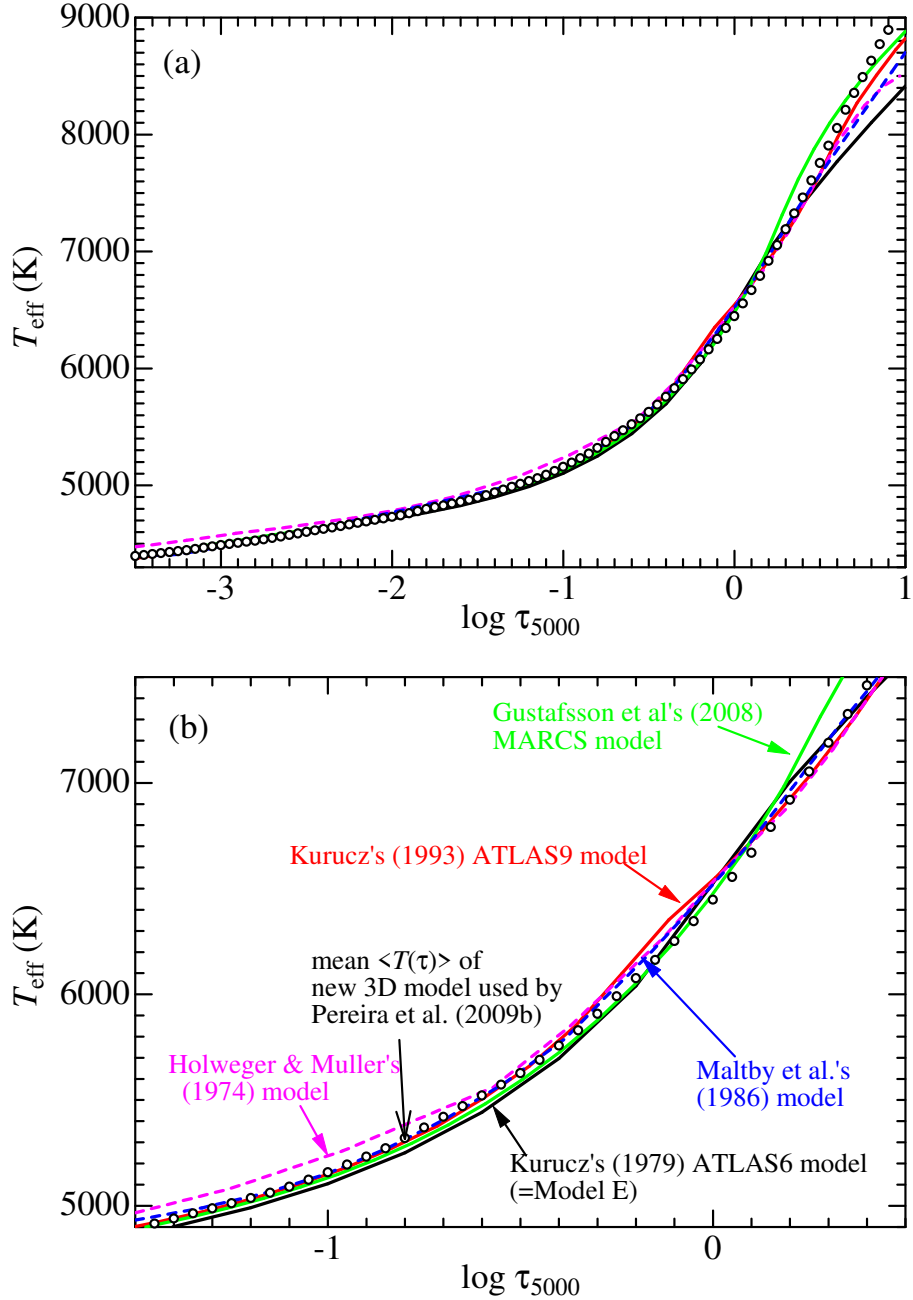}
    %%% \FigureFile(width,height){filename}
  \end{center}
\caption{
Comparison of the temperature structures of different solar photospheric 
models. The upper panel (a) shows the $T$ vs. $\tau_{5000}$ relation for the
$-3.5 \le \log\tau_{5000} \le +1.0$ region (wide view), while the lower
panel is for the $-1.5 \le \log\tau_{5000} \le +0.5$ region (zoomed view). 
Pink dashed line $\cdots$ Holweger and M\"{u}ller's (1974) model,
black solid line $\cdots$ Kurucz's (1979) ATLAS6 model (on which our Models E and C are based),
blue dashed line $\cdots$ Maltby et al.'s (1986) model,
red solid line $\cdots$ Kurucz's (1993) ATLAS9 model (with convective overshooting),
light-green solid line $\cdots$ Gustafsson et al.'s (2008) MARCS model, and 
open circles $\cdots$ mean $\langle T(\tau_{5000})\rangle$ of the new 3D model used by 
Pereira et al. (2009b)(read from figure 3 of their paper).
}
\end{figure}

\end{document}